\pdfoutput=1
\documentclass[12pt]{article}
\usepackage{graphicx}
\usepackage{amssymb,amsmath,amsfonts,palatino,amsthm}
\usepackage{amssymb}
\usepackage{epstopdf}
\newcommand{\beq}{\begin{equation}}
\newcommand{\eeq}{\end{equation}}

\newcommand{\f}{\begin{equation}}
\newcommand{\ff}{\end{equation}}

\newtheorem{theorem}{Theorem}

\newtheorem{case}{Case}

\newtheorem{definition}{Definition}
\begin{document}

\title{On Braid Excitations in Quantum Gravity}

\author{
Yidun Wan\thanks{Email address: ywan@perimeterinstitute.ca}
\\
\\
\\
Perimeter Institute for Theoretical Physics,\\
31 Caroline st. N., Waterloo, Ontario N2L 2Y5, Canada, and \\
Department of Physics, University of Waterloo,\\
Waterloo, Ontario N2J 2W9, Canada\\}
\date{October 5, 2007}
\maketitle

\vfill

\begin{abstract}
We propose a new notation for the states in some models of quantum
gravity, namely 4-valent spin networks embedded in a topological
three manifold. With the help of this notation, equivalence moves,
namely translations and rotations, can be defined, which relate the
projections of diffeomorphic embeddings of a spin network. Certain
types of topological structures, viz 3-strand braids as local
excitations of embedded spin networks, are defined and classified by
means of the equivalence moves. This paper formulates a mathematical
approach to the further research of particle-like excitations in
quantum gravity.
\end{abstract}
\vfill
\newpage
\tableofcontents
\newpage

\section{Introduction}

Loop quantum gravity had never been considered a candidate of the
unification of matter and gravity until a remarkable series of
discoveries emerged recently. First, Markopoulou and
Kribs\cite{Kribs2005} discovered that loop quantum gravity and many
related theories of dynamical quantum geometry have emergent
excitations which carry conserved quantum numbers not associated
with geometry or the gravitational field.

Around the same time, Bilson-Thompson\cite{Bilson-Thompson2005}
found that a composite or \textquotedblleft preon" model of the
quarks, leptons and vector bosons could be coded in the possible
ways that three ribbons can be braided and twisted. This suggested
that the particles of the standard model could be discovered amidst
the emergent braid states and their conserved quantum numbers
associated with those of the standard model. One realization of this
was then given in \cite{Bilson-Thompson2006}, for a particular class
of dynamic quantum geometry models based on 3-valent quantum
spin-networks obtained by gluing trinions together. These are coded
in the knotting and braiding of the edges of the spin network; they
are degrees of freedom because of the basic result that quantum
gravity or the quantization of any diffeomorphism invariant gauge
theory has a basis of states given by embeddings up to
diffeomorphisms of a set of labeled graphs in a spatial manifold.
Indeed, the role of the braiding of the edges of the graphs had been
a mystery for many years.

However, spin foam models in $3+1$ dimensions involve embedded
4-valent spin networks\cite{Markopoulou1997}. It is then natural to
ask if there are conservation laws associated with braids in
4-valent spin-networks. Besides, quantum gravity with a positive
cosmological cons- tant\cite{Smolin2002} and quantum deformation of
quantum gravity\cite{Major1995} suggest the framing of embedded
spin-networks. In this paper we extend the investigation of the
braid excitations from the 3-valent case to the 4-valent case. We
study (framed) 4-valent spin-networks embedded in 3D.

Due to the complexity of embedded 4-valent spin-networks, to deal
with the braid excitations of them we need a consistent and
convenient mathematical formalism. In this paper, which is the first
of a series of papers on the subject, we first propose a new
notation of the embedded (framed) 4-valent spin-networks and define
what we mean by braids, then discuss equivalence moves with the help
of our notation, which relate all diffeomorphic embedded 4-valent
graphs and form the graphical calculus of the kinematics of these
graphs, and at the end present a classification of the braids. These
results are key to our subsequent papers. We focus on 3-strand
braids, which are the simplest non-trivial and interesting braid
excitations living on embedded 4-valent spin-networks.

\bigskip

\section{Notation}

\begin{figure}[h]
\begin{center}
\includegraphics[
natheight=1.434700in,
natwidth=2.948200in,
height=1.4702in,
width=2.9914in
]{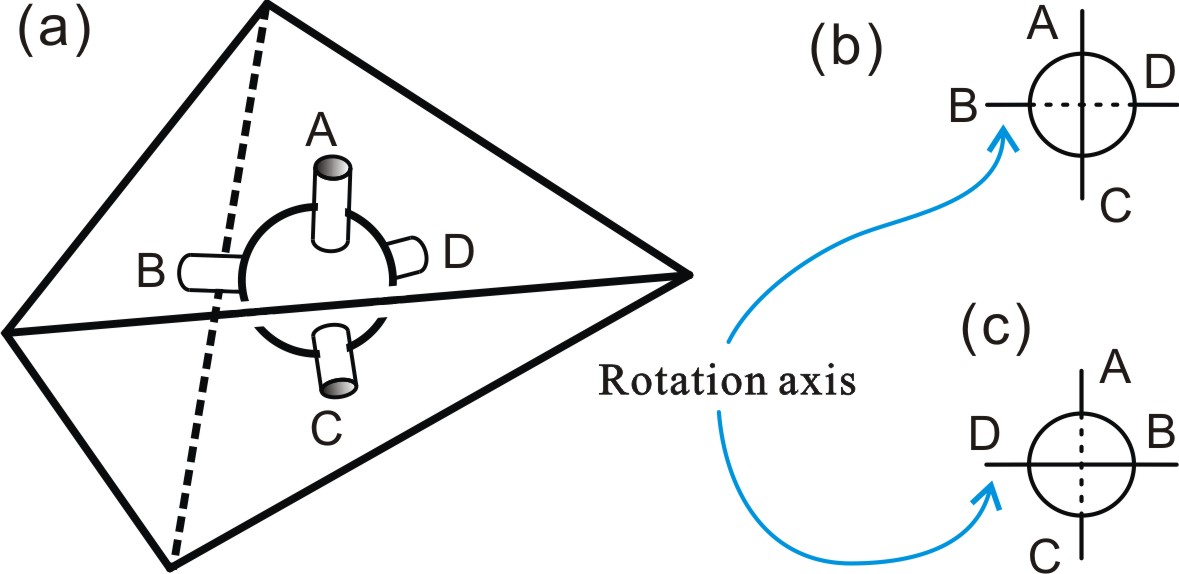}
\end{center}
\caption{(a) is a tetrahedron and its dual node. There are two
orientations of a node that can appear in a diagram, (b) denotes the
$\oplus$ state, while (c) denotes the $\ominus$ state.}
\label{notation}
\end{figure}

Firstly, we fix the notation, namely a tube-sphere notation. We work
in the category of framed graphs, in particular the two dimentional
projections representing embedded framed 4-valent spin networks up
to diffeomorphisms. There is a single diffeomorphism class of nodes.
We therefore represent nodes by rigid 2-spheres and edges by tubes.
Such a node can be considered locally dual to a tetrahedron, as
shown in Fig. \ref{notation}(a). If the spin-nets are not framed, we
simply reduce tubes to lines but still keep spheres as nodes. To
fully characterize the embedding of a spin-net in a 3-manifold, we
assume that not only the nodes are rigid, i.e. they can only be
rotated or translated, but also the positions on the node where the
edges are attached are fixed. This requirement and the local duality
ensures the non-degeneracy of the nodes, i.e. no more than two edges
of a node are co-planar. For the convenience of calculation, we
simplify the tube-sphere notation in Fig. \ref{notation}(a) to Fig.
\ref{notation}(b), in which 1) the sphere is replaced by a solid
circle; 2) the two tubes in the front, $A$ and $C$ in (a), are
replaced by a solid line piercing through the circle in (b); and 3)
the two tubes in the back, $B$ and $D$ in (a) are substituted by $B$
and $D$ in (b) with a dashed line connecting them through the
circle. There is no loss of generality in taking this simplified
notation, because one can always arrange a node in the two states
like Fig. \ref{notation}(b) \& (c) by diffeomorphism before taking a
projection. Due to the local duality between a node and a
tetrahedron and the fact that all the four edges of a node are on an
equal footing, if we choose one of the four edges of a node at a
time, the other three edges are still on an equal footing, in
respect to a rotation symmetry with the specially chosen edge as the
rotation axis, e.g. the edge $B$ in Fig. \ref{notation}(b) \& (c).
This rotation symmetry will be discussed in detail in the next section.%

\begin{figure}[p]
\begin{center}
\includegraphics[
natheight=1.577400in, natwidth=2.945600in, height=1.6137in,
width=2.9888in ]{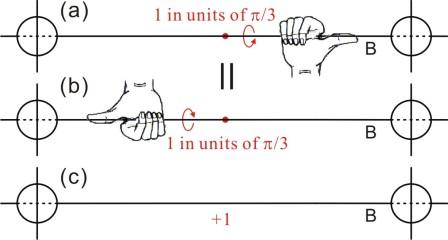}
\end{center}
\caption{The 1 unit of twist (equivalent to a $\pi/3$ rotation) in
(a) means cut to the right of the red dot and twist as shown. This
is is equivalent to the opposite twist on the opposite side of the
red dot, as shown in (b). Thus, both may be represented as in (c),
by a label of $+1$ of edge $B$.}
\label{notationtwist}
\end{figure}

There could exist twists on embedded tubes, e.g. the $\pi/3$-twist
on the edge $B$ with respect to the solid red dot, shown in Fig.
\ref{notationtwist}(a). Note that we put twists in the unit of
$\pi/3$ for two reasons. The first reason is that the possible
states by which a node may be represented in a projection can be
taken into each other by $\pi/3$ rotations around one of the edges
of the node (this will become clear in Section \ref{subsecRot}). By
the local duality of a node to a tetrahedron, these correspond to
the $\pi/3$ rotations that relate the different ways that two
tetrahedra may be glued together on a triangular face. These
rotations create twists in the edges and, as a result of the
restriction on projections of nodes we impose, set the twists in a
projection of an edge of a spin network in units of $\pi/3$. The
other reason is that the least twist distinguishable from zero of a
piece of tube in a projection is $\pi/3$ and all higher twists
distinguishable from each other in the projection must then be
multiples of $\pi/3$.

Because an edge is always between two nodes and a rotation of a node
creates/annihilates twists on its edges, one usually needs to
specify the fixed point on an edge with respect to which a twist is
counted, as shown in Fig. \ref{notationtwist}(a). In this manner,
the 1 unit of twist in Fig. \ref{notationtwist}(a) is obviously
equivalent to that in \ref{notationtwist}(b), which is the same
amount of twist in the opposite direction on the other side of the
fixed point. Interestingly, both twists in Fig.
\ref{notationtwist}(a) and (b) are right-handed twists if one point
his/her right thumb to the node on the same sides of the fixed point
as that of the twists; therefore, we can unambiguously assign the
same value to them, namely $+1$ (unit of $\pi/3$). This provides a
way of simplifying the notation of twist, i.e. we can simply label
an edge with a (left-) right-handed twist\ a (negative) positive
integer. For example, Fig. \ref{notationtwist}(a) and (b) can be
replaced by
\ref{notationtwist}(c) without ambiguity.%

Recalling the rotation axis mentioned before, one can assign states
to a node with respect to its rotation axis. If the rotation axis is
an edge in the back, we say the node is in state $\oplus$, or is
simply called a $\oplus$-node, e.g. Fig. \ref{notation}(b) with edge
$B$ as the rotation axis. Otherwise, if the rotation axis is an edge
in the back, the node is called a $\ominus$-node or
in the state $\ominus$, e.g. Fig. \ref{notation}(c) with edge $D$.%

\subsection{Framed and unframed spin networks}
The results of this paper will refer to the case of framed spin
networks, defined above. However, unframed graphs are used in loop
quantum gravity and it is useful to have results then for that case
as well. The particular notation of unframed graphs is obtained from
the framed case discussed here by dropping information about twists
of the edges (which thus represent curves rather than tubes), but
keeping the nodes as rigid spheres, locally dual to tetrahedra. This
is necessary so that the evolution moves are well defined for
unframed embedded graphs, which will be explained in the second of
this series of papers.

In the rest of this paper we refer always to the framed case.
Results for the unframed case will be understood from those for the
framed case by neglecting the twists of the edges, unless we
explicitly describe them.

\begin{figure}[p]
\begin{center}
\includegraphics[
natheight=0.909800in, natwidth=2.764800in, height=0.9426in,
width=2.808in ]{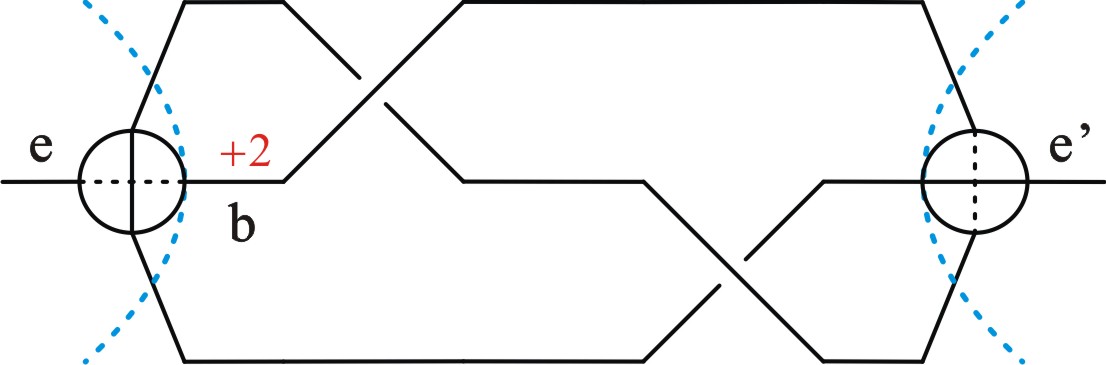}
\end{center}
\caption{A typical 3-strand braid formed by the three common edges
of two end-nodes. The region between the two dashed line satisfies
the definition of an ordinary braid. Edges $e$ and $e^{\prime}$ are
called external edges. There is also a right handed twist of 2 units
on strand $b$. In this figure the left handed node is in a $\oplus$
state while the right handed node is in a $\ominus$ state.}
\label{braid}
\end{figure}

\section{Braids}
Equipped with the notation defined above, we are interested in a
type of topological structures as sub-structures of embedded
4-valent spin networks, namely 3-strand braids, which are defined as
follows.

\begin{definition}
\label{defbraid}A \textbf{3-strand braid (}or a \textbf{braid} for
simplicity\textbf{)} is a sub-spinnet of an embedded 4-valent spin
network, which is a three dimensional object formed by two nodes
with three common edges, now named \textbf{strand}s; the two nodes
are called \textbf{end-node}s, each of which has one and only one
free edge, called an \textbf{external edge}. The two dimensional
projections of these braids denoted in our notation are called braid
diagrams, a typical example of which is shown in Fig. \ref{braid}.
The following conditions should be satisfied:

\begin{enumerate}
\item if braids are arranged horizontally, then the (left) right external edge
of a braid can always be the (left-) right-most edge of the (left)
right end-node, and always stretches to the (left) right, which has
no tangles with the strands, for the left part of the braid diagram
in Fig. \ref{notbraid}(a) as an example;

\item what is captured between the two end-nodes, e.g. the region between the
two dashed lines in Fig. \ref{braid}, should meet the definition of
braid in the ordinary braid theory, for the braid diagram in Fig.
\ref{notbraid}(b) as an example;

\item the three strands of a braid are never tangled with any other edge of
the spin-net, as illustrated in right side of the braid diagram in
Fig. \ref{notbraid}(a), for example.%

\begin{figure}
[p]
\begin{center}
\includegraphics[
natheight=1.909500in, natwidth=2.772600in, height=1.9476in,
width=2.815in ] {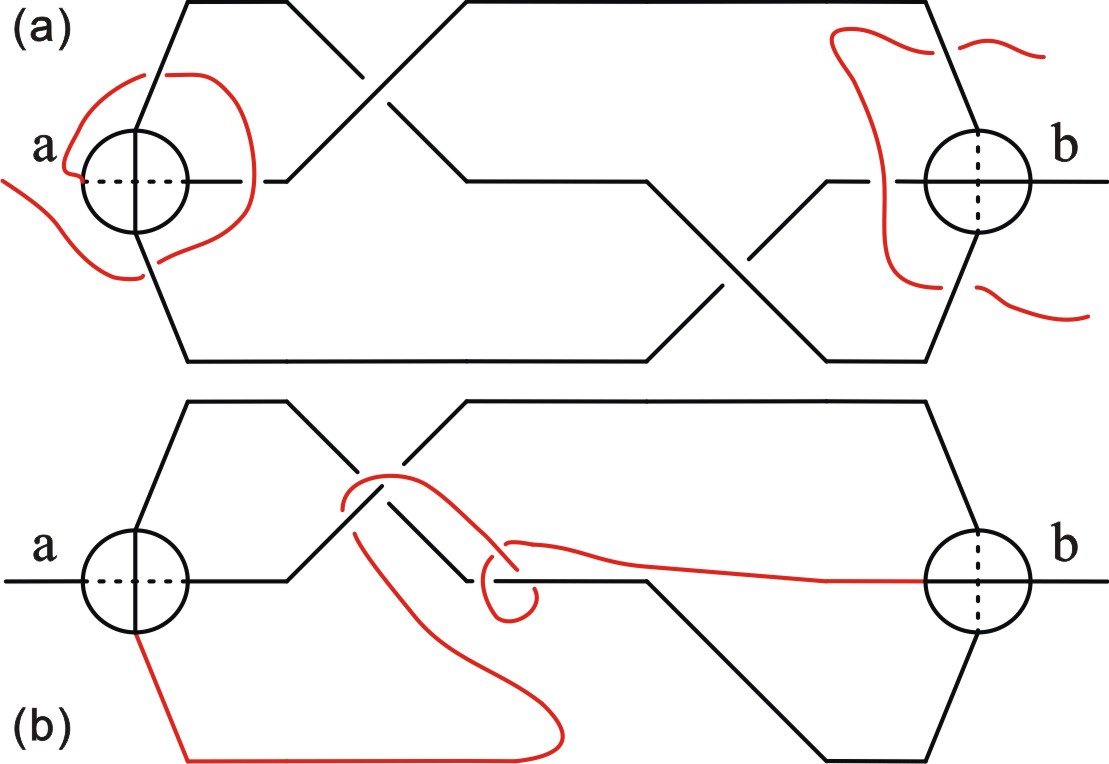} \caption{(a) is not a braid due to
the tangle between the external edge $a$ and the common edges of the
two nodes, and/or the tangle between the common edges and another
edge connecting elsewhere in the whole spinnet. (b) is not a braid
either because the region captured between the two nodes does not
satisfy the ordinary definition of a braid.} \label{notbraid}
\end{center}
\end{figure}

\end{enumerate}
\end{definition}

We would like to emphasize that the braids defined above are 3D
structures, each of which has many diffeomorphic embeddings that are
represented by their 2D projections in our notation. As a result,
the 2D projections of many braids, i.e. their braid diagrams, which
appear to be different are actually equivalent to each other in the
sense of diffeomorphism. The precise set of equivalence relations
will be the topic of the next section. Bearing this in mind, in the
rest of the paper we are not going to distinguish braids from their
braid diagrams, unless an ambiguity arises.

These kinds of braids are different from the braids in the context
of ordinary braid theory, since the two end-nodes of such a braid
are topologically significant to the state of the braid. These braid
are stable under a certain stability condition regarding the
evolution of spin-nets, which will be brought up in the companion
paper. However, in this paper, we focus only on the intrinsic
properties of these braids, or in other words the pure topological
properties of the braids up to diffeomorphism, i.e. without dynamic
evolution. To do so, we need to first describe the non-dynamical
operations that can be applied to the embedded 4-valent spin
networks.

\begin{figure}
[p]
\begin{center}
\includegraphics[
natheight=0.460100in, natwidth=2.940400in, height=0.4903in,
width=2.9836in ] {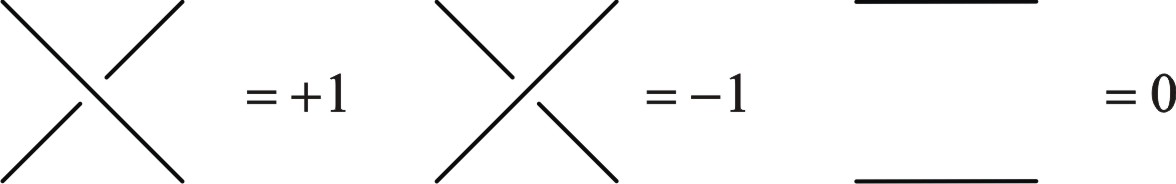} \caption{The assignments of
right-handed crossing, left-handed crossing, and null crossing
respectively from left to right.} \label{assignment}
\end{center}
\end{figure}

We can assign a number to a crossing according to its chirality, viz
$+1$ for a right-handed crossing, $-1$ for a left-handed crossing,
and $0$ otherwise.
Fig. \ref{assignment}\ shows this assignment.%

Such a scheme of assignment will become useful in the subsequent
discussions.

\bigskip

\section{Equivalence Moves}

As aforementioned, the tube diagrams of an embedded spin network
belong to different equivalence classes. It is therefore obligatory
to characterize these equivalence classes by equivalence relations.
To do so, one needs to find the full set of local moves, operating
on the nodes and edges, which don't change the diffeomorphism class
of the embedding of a diagram. In the discussion below we work in
the framed case. In the unframed case, one just ignores the twists.

An obvious set of equivalence moves consists of the usual three
Reidemeister moves, framed or unframed, whose details are not
repeated here; these moves will be applied without further notice.
More importantly, there are two kinds of equivalence moves that can
be peculiarly defined on an embedded 4-valent spin-net, under which
two diagrams, in particular two braids, that are related by a
sequence of equivalence moves are thought to be equivalent. The
first kind composes of translation moves. The second type of
equivalence moves are rotations defined on the nodes.

\subsection{Translation Moves}

\begin{figure}
[p]
\begin{center}
\includegraphics[
natheight=1.241900in, natwidth=2.952500in, height=1.2773in,
width=2.9966in ] {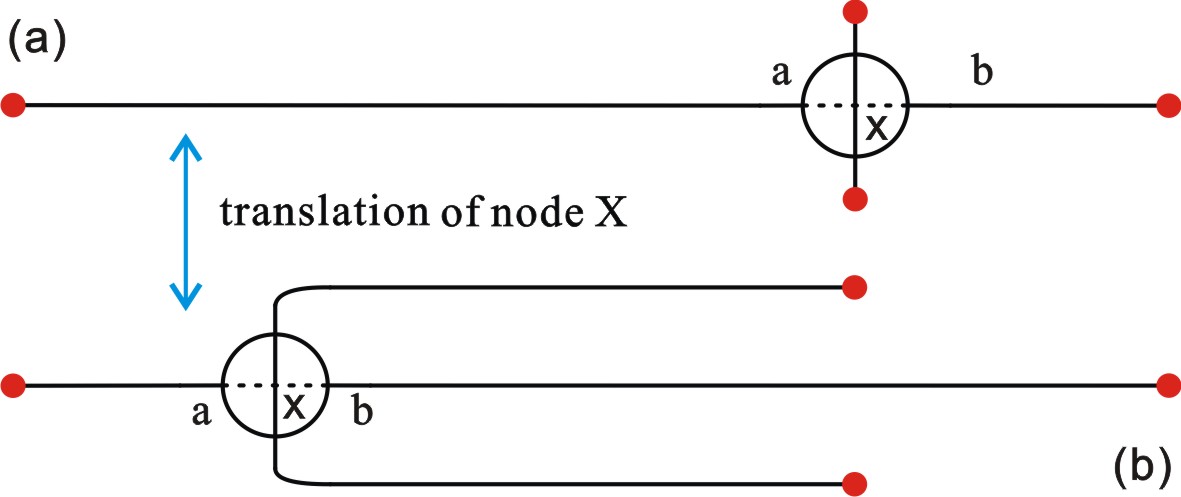} \caption{Red points I and J
represent other nodes where edges $a$ and $b$ are attached to. (b)
is obtained from (a)\ by translating node $X$ from right to left,
and vice versa.} \label{translation}
\end{center}
\end{figure}

\begin{figure}
[p]
\begin{center}
\includegraphics[
natheight=0.952200in, natwidth=2.948200in, height=0.985in,
width=2.9914in ] {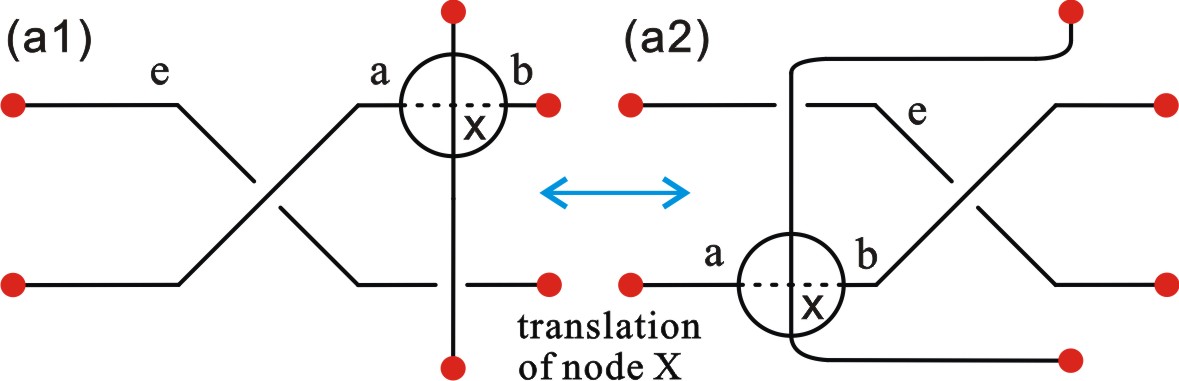} \caption{Red points represent
other nodes to which edges $a$ and $b$ are attached. (a1) and (a2)
can be transformed into each other by translating node $X$. To
transform (b1) into (b2) is not allowed due to the tangle produced
by translating node $X$.} \label{translationX}
\end{center}
\end{figure}

We discuss translation moves first. Translation moves, which are in
fact extended Reidemeister type moves, involve not only the edges
but also the nodes of an embedded spin-net; they reflect the
translation symmetry of the embedded spin-nets. Let us look at the
simplest example first. Fig. \ref{translation}(a) shows a node $X$
connected to other places of the network via its four edges; red
points represent attached points on other nodes. One can slide the
node $X$ along its edge $a$ to the left, which leads to Fig.
\ref{translation}(b); this does not change anything of the topology
of the embedded spin-net. Fig. \ref{translationX} illustrates more
complicated cases where a crossing is taken into account.

In Fig. \ref{translationX}(a1) there is a node $X$ and a crossing;
however, since the crossing is between the edge $a$ of node $X$ and
the edge $e$ of some other node and node $X$ together with all its
edges are above edge $e$, one can safely translate node $X$ along
edge $a$ to the left passing the crossing, which results in Fig.
\ref{translationX}(a2), in which the crossing turns out to be
between edge $e$ and edge $b$. This, which is obviously a
symmetry, may be understood as a Reidemeister move II.%

\subsection{Rotations}
\label{subsecRot}

Apart from the translation symmetry, there is also a rotation
symmetry that gives rise to rotations defined on a node, with
respect to one of its four edges, of an embedded spin-net. These
rotations are not those with rigid metric but only the ones that
change projections of an embedding, without
affecting diffeomorphisms.%

\subsubsection{$\pi/3$-Rotations}

\begin{figure}
[p]
\begin{center}
\includegraphics[
natheight=2.082500in, natwidth=2.943000in, height=2.1101in,
width=3.5699in ] {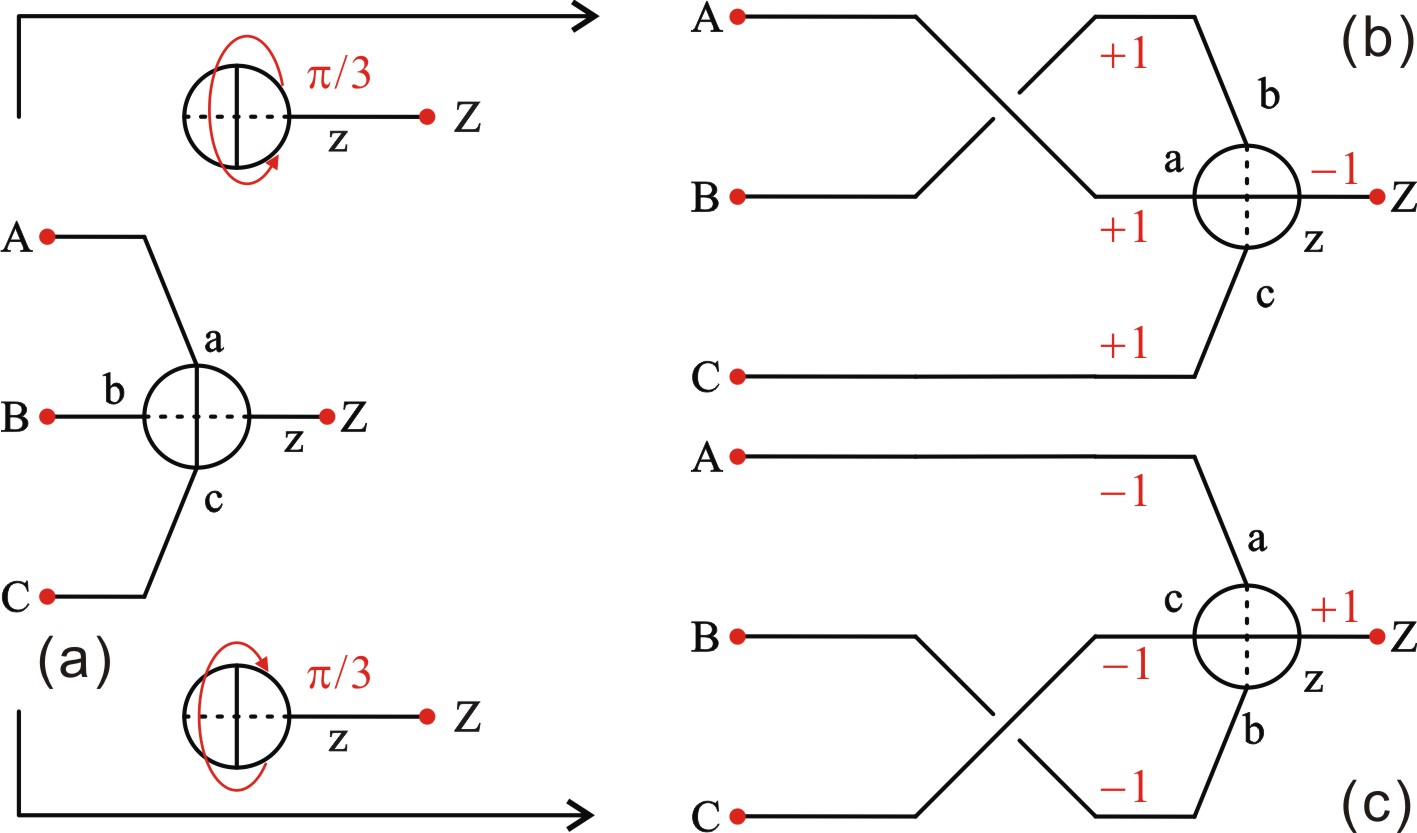} \caption{(b) \& (c) are results of
(a) by rotating the $\oplus$-node in (a) w.r.t. edge $z$ in two
directions respectively. Points $A$, $B$, $C$, and $Z$ are assumed
to be connected somewhere else and are kept fixed during the
rotation. All edges of the node gain the same amount of twist after
rotation.} \label{pi3rot+}
\end{center}
\end{figure}

\begin{figure}
[p]
\begin{center}
\includegraphics[
natheight=2.147300in, natwidth=2.943000in, height=2.1223in,
width=3.5699in ] {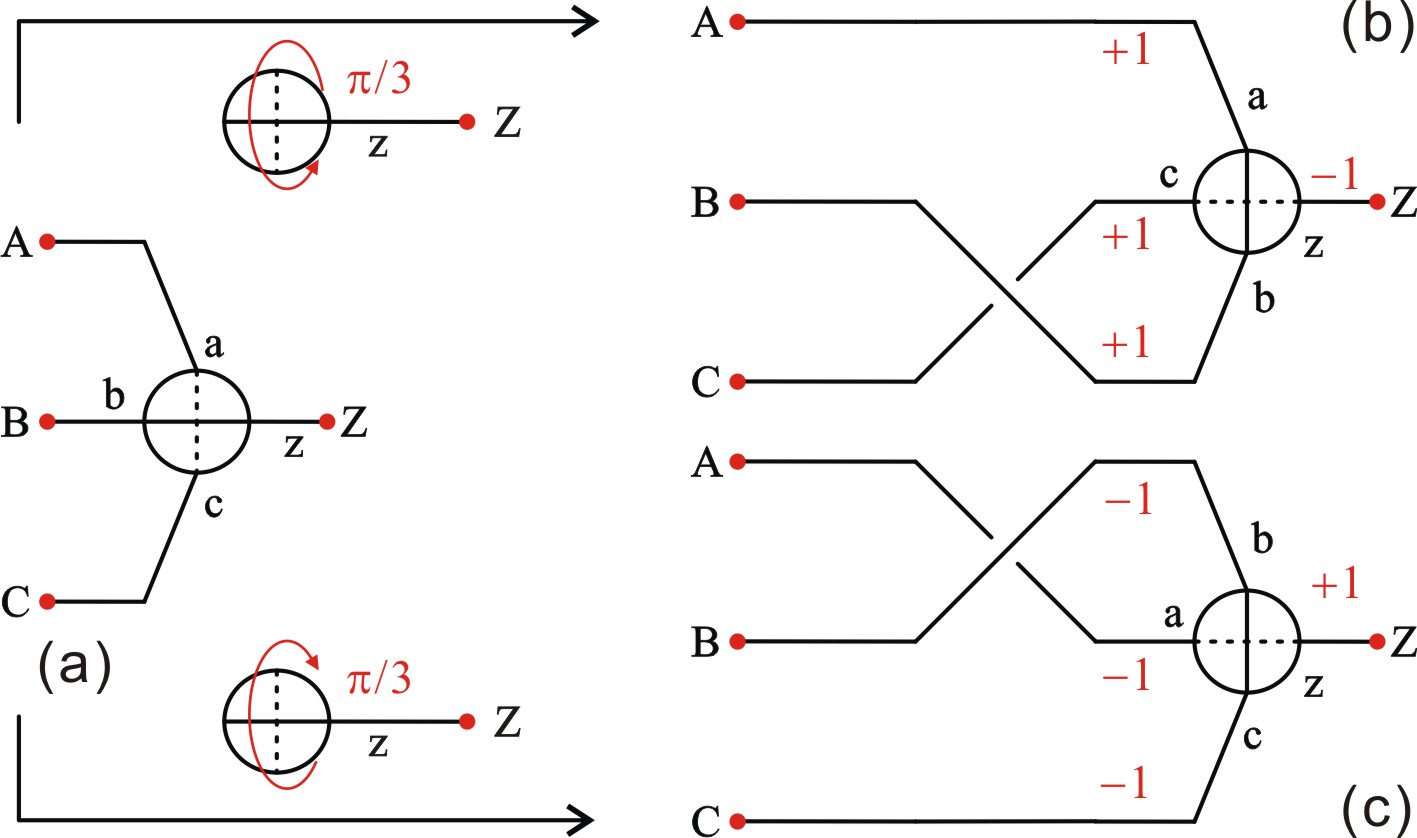} \caption{(b) \& (c) are results of
(a) by rotating the $\ominus$-node in (a) w.r.t. edge $z$ in two
directions respectively. Points $A$, $B$, $C$, and $Z$ are assumed
to be connected somewhere else and are kept fixed during the
rotation. All edges of the node gain the same amount of twist after
rotation.} \label{pi3rot-}
\end{center}
\end{figure}

As mentioned before, $\pi/3$ rotations take states representing a
node in a projection into each other. It is time to see in detail
how these rotations affect a subgraph consisting of a node and its
four edges. Fig. \ref{pi3rot+} shows such a rotation in the case
where the node is in a $\oplus$-state with respect to the chosen
rotation axis before imposing the rotation, while Fig. \ref{pi3rot-}
illustrates the opposite case.

A $\pi/3$ rotation creates a crossing of two edges of the node and
causes twists, which are explicitly labeled, on all edges of the
node. The twist number on the rotation axis of a node is always
opposite to that of the other three edges of the node. Note that a
$\pi/3$-rotation changes the state of a node, as shown in Figures
\ref{pi3rot+} and \ref{pi3rot-}, i.e. if the node is in state
$\oplus$ before the rotation, it becomes a $\ominus$-node after the
rotation. This is the key to the first reason that we put the twists
in an edge in units of $\pi/3$. A $\pi/3$ rotation relates two
projections of an embedded spin network, which belong to the same
diffeomorphism class.

\subsubsection{$2\pi/3$-Rotations}

\begin{figure}
[p]
\begin{center}
\includegraphics[
trim=0.000000in 0.000000in -0.004709in -0.001288in,
natheight=2.147300in, natwidth=2.943000in, height=5.3949cm,
width=9.0808cm ] {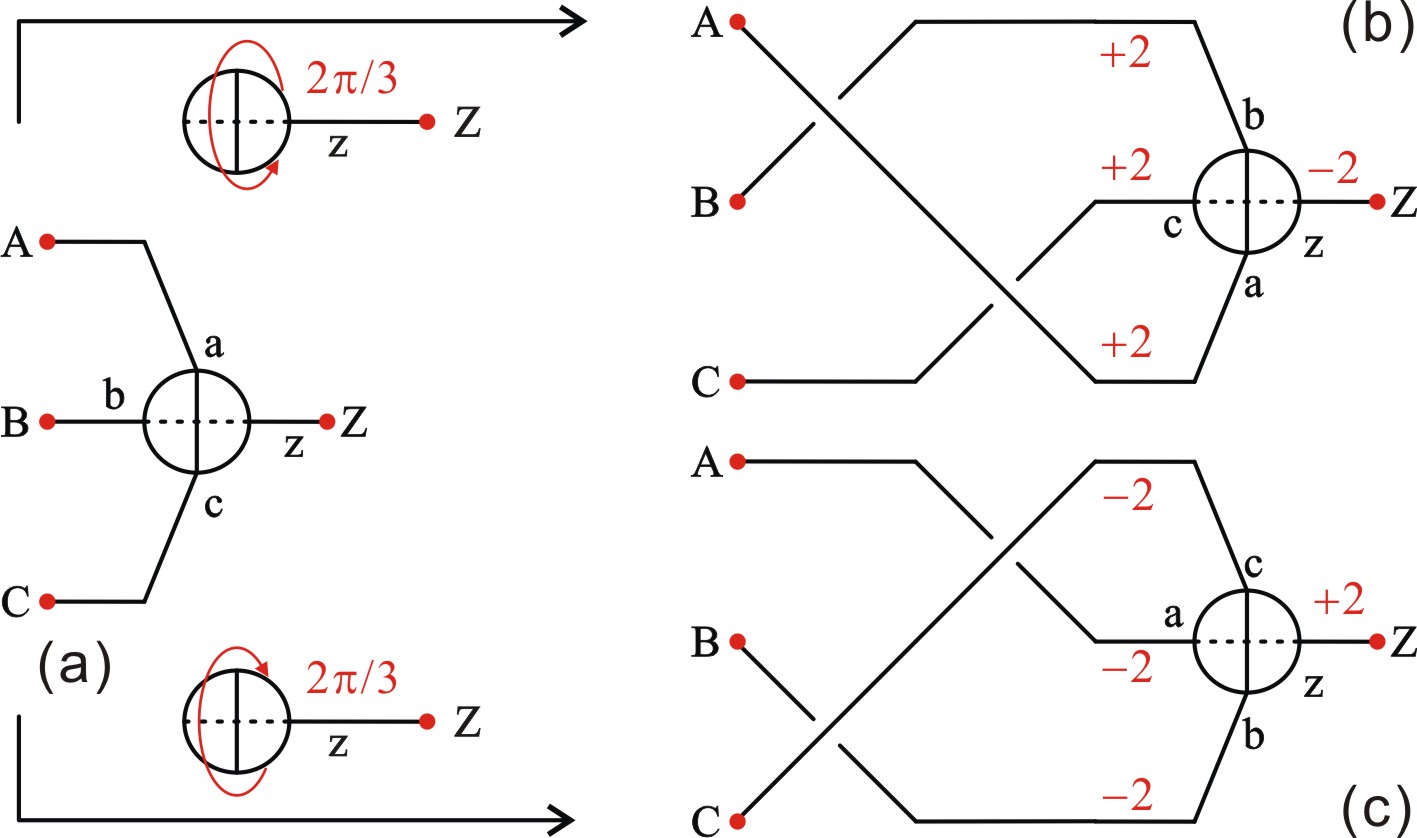} \caption{(b) \& (c) are results of
(a) by rotating the $\oplus$-node in (a) w.r.t. edge $z$ in two
directions respectively. Points $A$, $B$, $C$, and $Z$ are assumed
to be connected somewhere else and are kept fixed during the
rotation. All edges of the node gain the same amount of twist after
rotation.} \label{2pi3rot+}
\end{center}
\end{figure}

\begin{figure}
[p]
\begin{center}
\includegraphics[
natheight=2.082500in, natwidth=2.943000in, height=2.1101in,
width=3.5699in ] {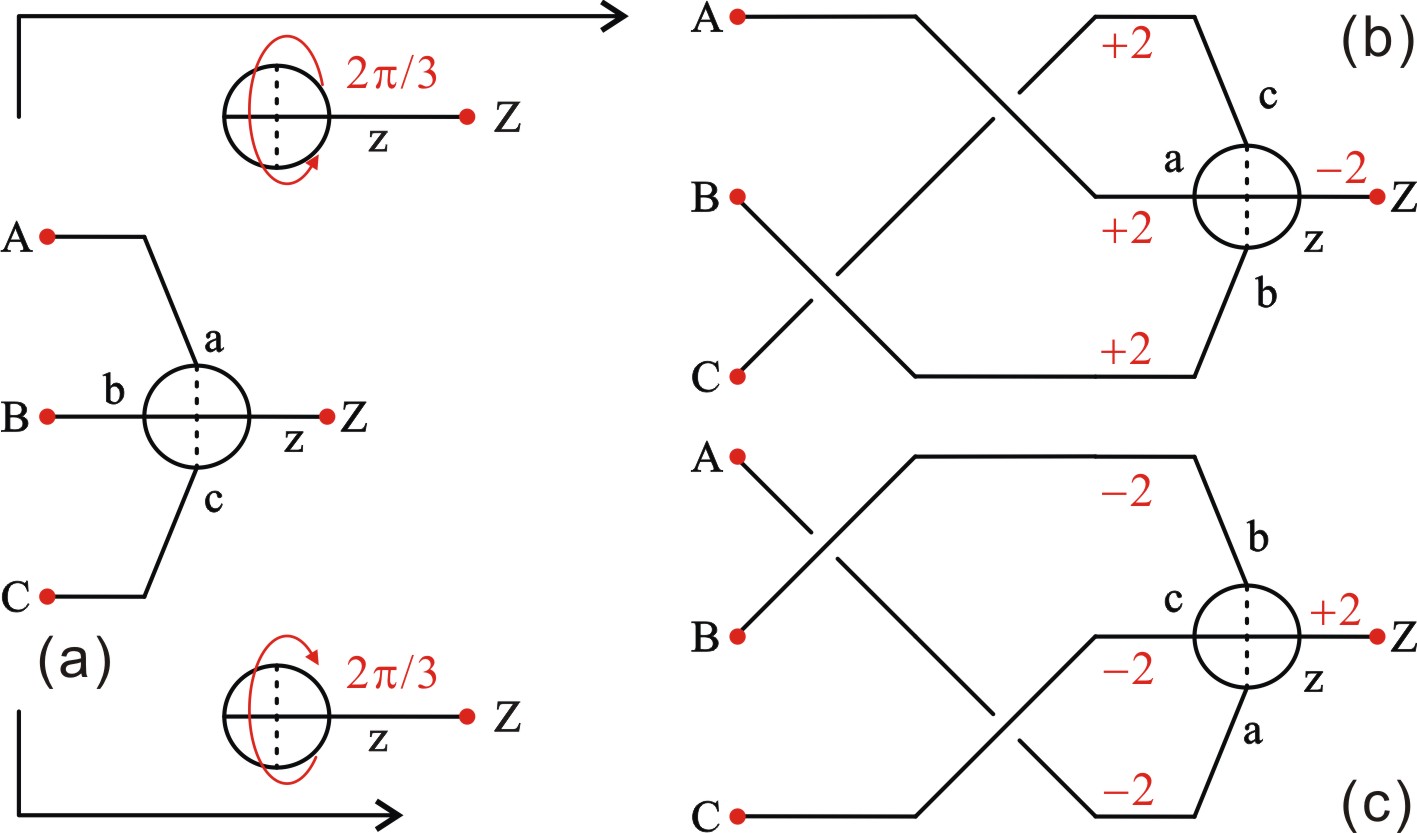} \caption{(b) \& (c) are results of
(a) by rotating the $\ominus$-node in (a) w.r.t. edge $z$ in two
directions respectively. Points $A$, $B$, $C$, and $Z$ are assumed
to be connected somewhere else and are kept fixed during the
rotation. All edges of the node gain the same amount of twist after
rotation.} \label{2pi3rot-}
\end{center}
\end{figure}

Two consecutive $\pi/3$ rotations certainly give rise to a $2\pi/3$
rotation. However, it is intuitive to understand $2\pi/3$ rotations
in a more topological way. Obviously, rotating a tetrahedron by an
angle of $2\pi/3$ with respect to the normal of any of the four
faces of the tetrahedron does not change the view of it. Therefore,
by the local duality between a node and a tetrahedron, as long as an
edge of a node is chosen, the other three edges of the node are on
an equal footing. If we rotate a node with respect to any of its
four edges by $2\pi/3$, the resulting diagram should be
diffeomorphic to, or in our context equivalent to, the original one.
In Fig. \ref{2pi3rot+} and Fig. \ref{2pi3rot-} we list all the
$2\pi/3$-rotations.

Each of such rotations generate two crossings and twists on all four
edges. The twist number on the rotation axis of a node is always
opposite to that of the other three edges of the node. Note that a
$2\pi/3$ rotation does not change the state of a node with respect
to the rotation axis, i.e. if a node is in state $\oplus$ with
respect to $z$ before the rotation, it is still a $\oplus$-node
after the rotation.

\subsubsection{$\pi$-Rotations}

\begin{figure}
[p]
\begin{center}
\includegraphics[
natheight=2.247700in, natwidth=2.943000in, height=2.2226in,
width=3.5699in ] {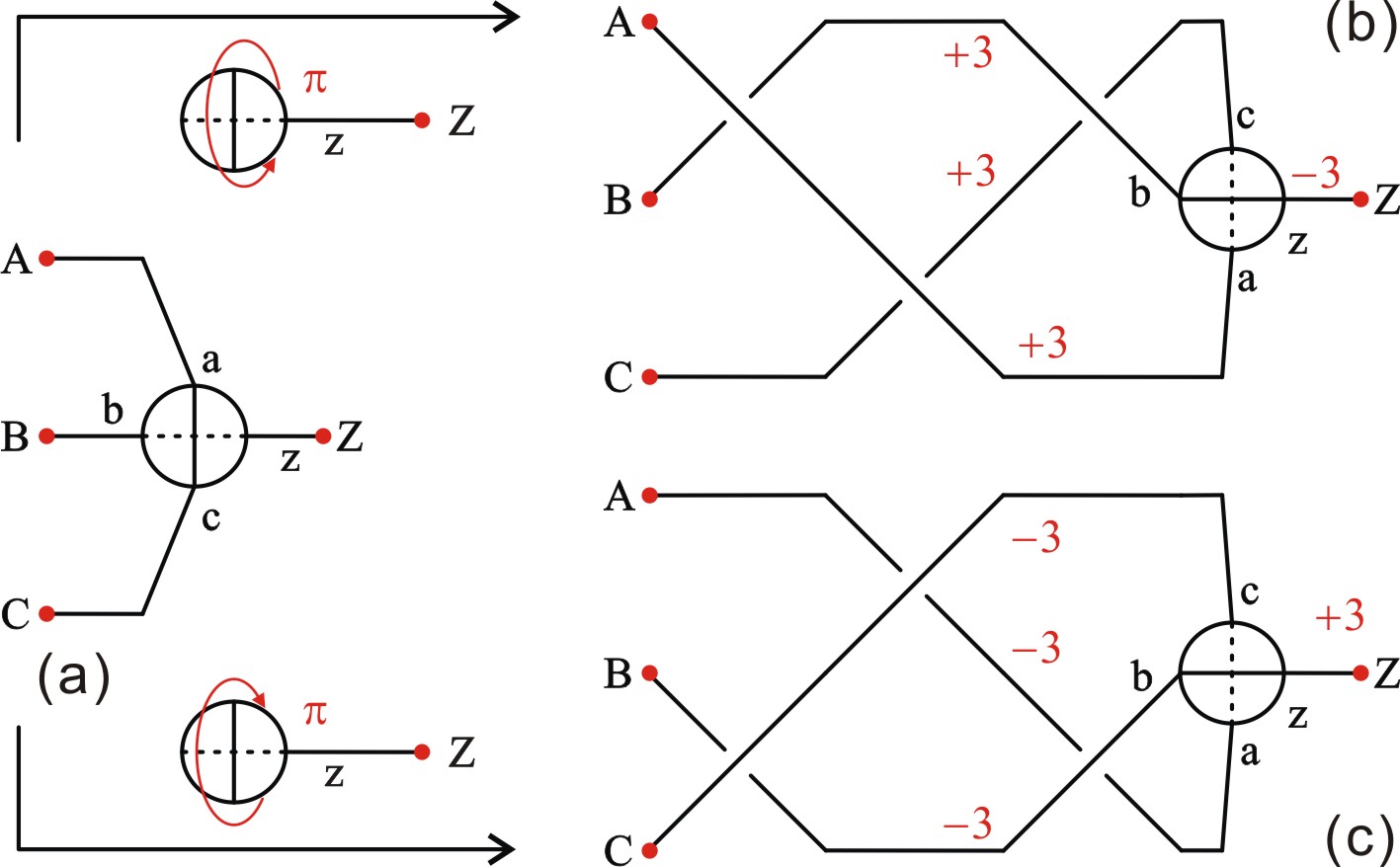} \caption{(b) \& (c) are results of (a)
by rotating the $\oplus$-node in (a) w.r.t. edge $z$ in two
directions respectively. Points $A$, $B$, $C$, and $Z$ are assumed
to be connected somewhere else and are kept fixed during the
rotation. All edges of the node gain the same amount of twist after
rotation.} \label{pirot+}
\end{center}
\end{figure}

\begin{figure}
[p]
\begin{center}
\includegraphics[
natheight=2.247700in, natwidth=2.943000in, height=2.2226in,
width=3.5699in ] {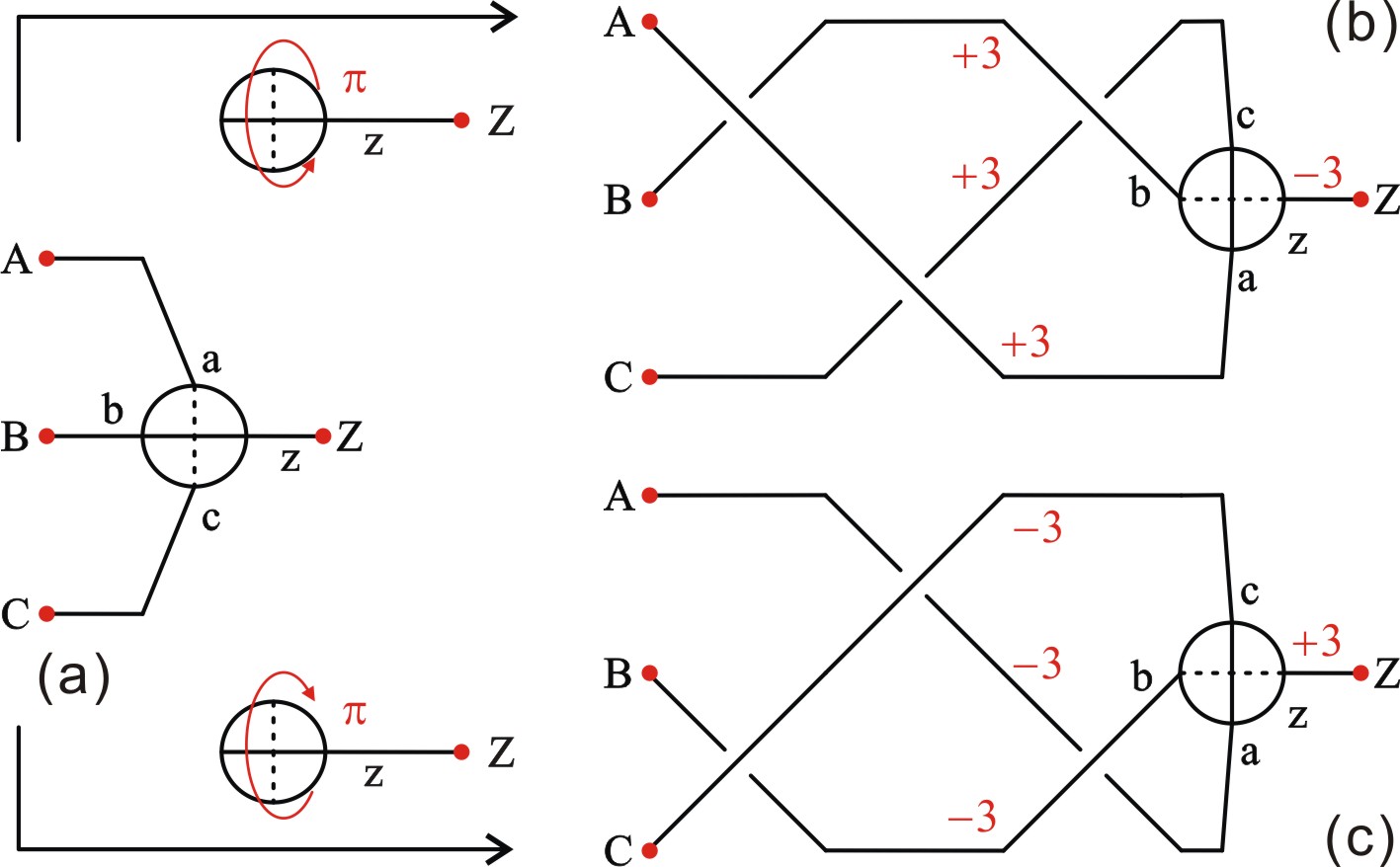} \caption{(b) \& (c) are results of (a)
by rotating the $\ominus$-node in (a) w.r.t. edge $z$ in two
directions respectively. Points $A$, $B$, $C$, and $Z$ are assumed
to be connected somewhere else and are kept fixed during the
rotation. All edges of the node gain the same amount of twist after
rotation.} \label{pirot-}
\end{center}
\end{figure}

The $\pi/3$ and $2\pi/3$ rotations can be used to construct larger
rotations, for example the $\pi$ rotations, which also certainly do
not change the diffeomorphism class a projection belongs to. For the
convenience of future use, we depict these four possible
rotations in Fig. \ref{pirot+}%
\ and Fig. \ref{pirot-}.%

Note that a $\pi$ rotation changes the state of a node, i.e. if a
node is in state $\oplus$ with respect to its rotation axis before
the rotation, it becomes a $\ominus$-node with respect to the same
axis after the rotation.

$\pi/3$ rotations are the smallest building blocks of all possible
rotations; they are thus the generators of all rotations. This is
illustrated in Fig. \ref{pi3rot+} through Fig. \ref{pirot-}, each of
which can be directly used in a graphic calculation.

\begin{figure}
[ph]
\begin{center}
\includegraphics[
natheight=2.092800in, natwidth=2.932600in, height=2.1318in,
width=2.9767in ] {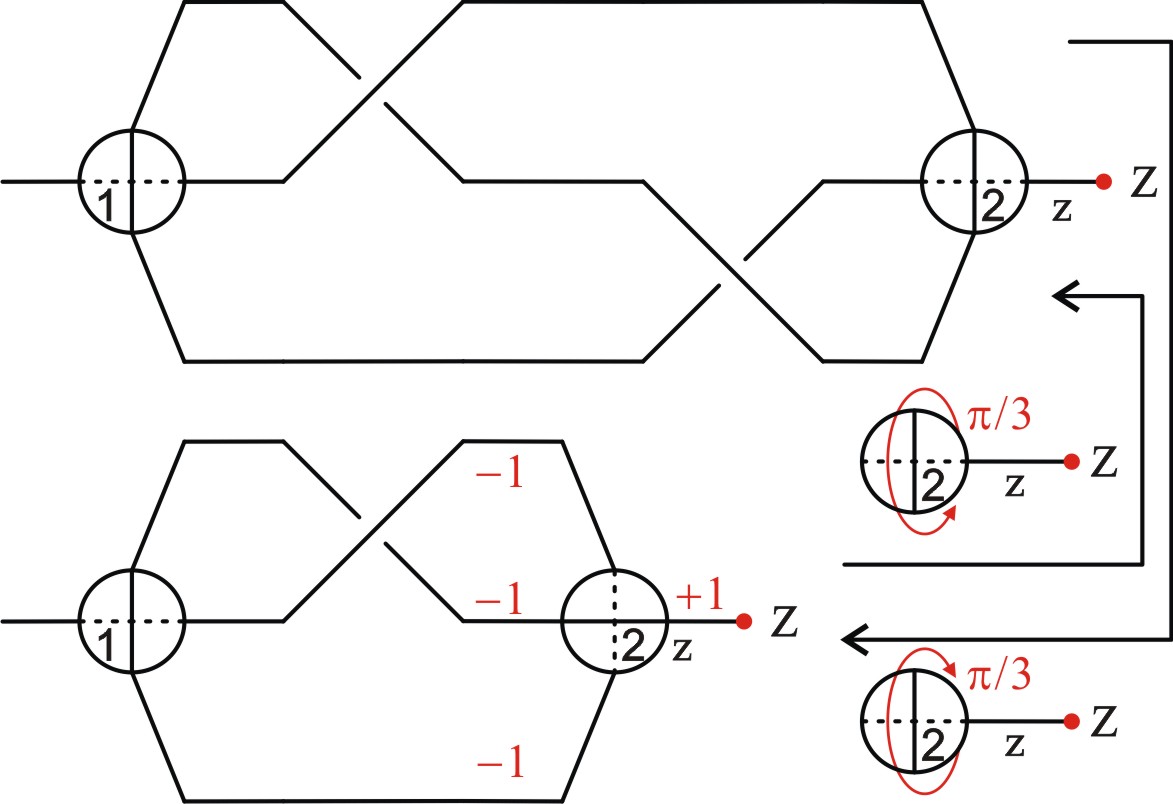} \caption{The two braids are
equivalent because they can can be transformed into each other by a
$\pi/3$-rotation of node 2.} \label{equibraids}
\end{center}
\end{figure}

Recall that all the equivalence moves defined above are
diffeomorphic operations on the embedded graphs. As an example, Fig.
\ref{equibraids} depicts two braids that can be deformed into each
other by a $\pi/3$ rotation of node 2 with respect
to its external edge $z$. We say these two braids are equivalent to each other.%

\begin{figure}
[ph]
\begin{center}
\includegraphics[
natheight=0.909800in, natwidth=2.772600in, height=0.9426in,
width=2.815in ] {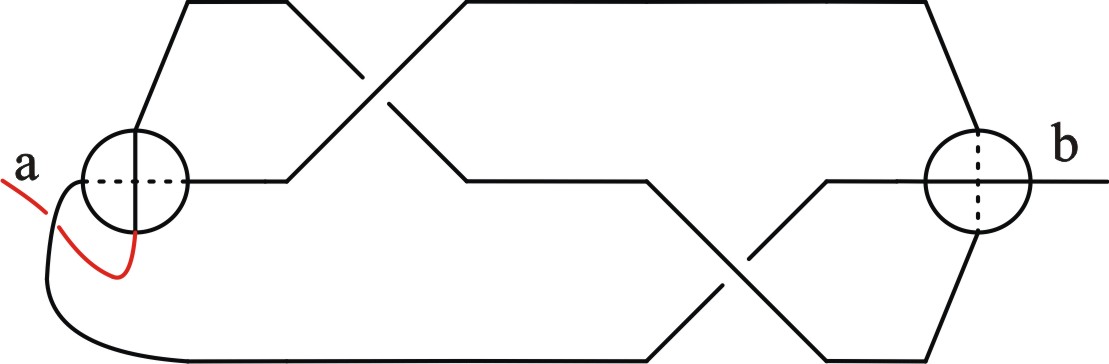} \caption{A sub-spinnet equivalent to
a braid by rotations with respect to the strands of the braid; it
does not satisfy our definition of a 3-strand braid any more.}
\label{badbraid}
\end{center}
\end{figure}

Note that for an end-node of a braid, only its external edge is
allowed to be the rotation axis, with respect to which the
equivalence rotation moves are applied. Otherwise, one may end up
with a situation similar to Fig. \ref{badbraid}, which does not
satisfy Definition \ref{defbraid}. Therefore, although sub-spinnets
like Fig. \ref{badbraid} are equivalent to well-defined braids by
rotation moves, they are not to be investigated because they
complicate the clear structure of braids and do not have any new
interesting property. Thus for simplicity we only allow the external
edge of an end-node of a braid to be the rotation axis. If a node is
not an end-node of a braid, any
of its four edges can be chosen as a rotation axis.%

By looking carefully at the rotations and the crossings and twists
generated accordingly one can find that the assignment of values to
crossings, shown in Fig. \ref{assignment}, is consistent with the
assignment of values to twists, shown in Fig. \ref{notationtwist}.

\subsection{Conserved quantity: effective twist number}

Given that the rotations and translations are well-defined
equivalence moves, there should be a conserved quantity, which is
the same before and after the moves. Rotations create or annihilate
twist and crossings simultaneously, we thus define a composite
quantity, christened \textbf{effective twist number of a rotation},
\begin{equation}
\Theta_{r}=\sum_{e=1}^{4}T_{e}-2\times\sum_{\substack{\text{all Xings}%
\\\text{created}}}X_{i},\label{thetar}%
\end{equation}
where $T_{e}$ is the twist number created by the rotation on an edge
of the node, $X_{i}$ is the crossing number of a crossing created by
the rotation between any two edges of the node, and the factor of 2
comes from the fact that a crossing always involve two edges. One
can easily check that the rotations in Fig. \ref{pi3rot+} through
Fig. \ref{pirot-} satisfy $\Theta_{r}\equiv0$. That is, rotations
have a zero effective twist number. Therefore we can enlarge
$\Theta_{r}$ to a more general quantity $\Theta_0$, the
\textbf{effective twist number of subdiagrams} of an embedded
spin-net, which are related by rotations of nodes, by taking into
account all the edges that are affected by rotations. We define
\begin{equation}
\Theta_0=\sum_{\substack{\text{all edges in a}\\\text{subdiagram}}}T_{e}%
-2\times\sum_{\substack{\text{all Xings in a}\\\text{subdiagram}}%
}X_{i},\label{theta0}%
\end{equation}

where $T_{e}$ is the twist number on an edge of the subdiagram,
$X_{i}$ is the crossing number of a crossing in the subdiagram.
Since $\Theta_{r}\equiv0$, $\Theta_0$ is indeed a conserved quantity
under rotations. Important examples of subdiagrams are braids, which
will become clear when we talk about propagation and interactions.

The effective twist number $\Theta_0$ in Eq. \ref{theta0} is also
found to be preserved by translation moves.

Note that the effective twist number is not defined in the unframed
case, simply because the unframed case has no notion of twists.

\bigskip

\section{Classification of Braids}

With the help of equivalence moves, in particular the rotation
moves, we can classify all possible 3-strand braids into two major
types, namely reducible braids and irreducible braids, whose
definitions are given below. The aforementioned restriction that
only the external edge of an end-node of a braid can be the rotation
axis of the node ensures the unambiguous assignments of states to
the end nodes of a braid and keeps the classification of braids
simple. Note that twists on edges are irrelevant to the calculation
in the section; they are thus neglected throughout the discussion.
Nevertheless, the results are valid for both framed and unframed
cases.

For the purpose of classifying the braids, we also consider braids
as if they are isolated regions in a graph.

\begin{definition}
\label{defRedubraid}A braid is called a \textbf{reducible braid}, if
it is equivalent to a braid with fewer crossings; otherwise, it is
\textbf{irreducible}.
\end{definition}

\begin{figure}
[ph]
\begin{center}
\includegraphics[
natheight=1.292900in, natwidth=2.945600in, height=1.3275in,
width=2.9888in ] {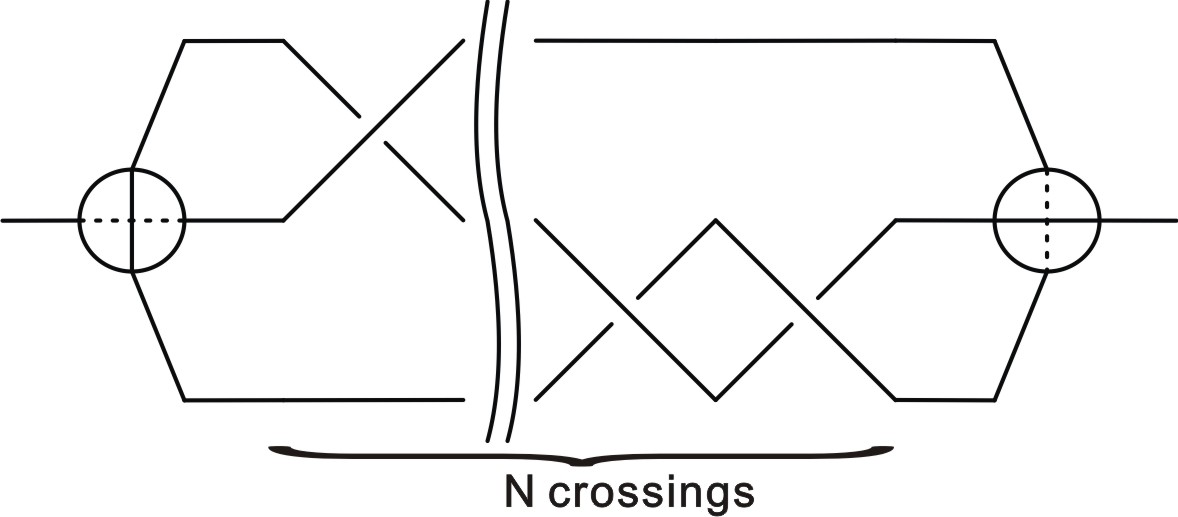} \caption{This braid can be
represented by $\oplus\left( \protect\begin{array} [c]{ccccc}
-1 & \cdots & \cdots & 0 & 0\protect\\
0 & \cdots & \cdots & +1 & +1 \protect\end{array} \right) \ominus$.}
\label{matrixbraid}
\end{center}
\end{figure}

The braid on the top part of Fig. \ref{equibraids} is an example of
a reducible braid, whereas the one at the bottom of the figure is an
irreducible braid. To classify the braids in a convenient way, we
need a new notation and some auxiliary definitions. Since we have a
way of assigning crossings integers $+1$ or $-1$, as in Fig.
\ref{assignment}, we can use $2\times N$ matrices with two end-nodes
in either state $\oplus$ or $\ominus$ to denote a 3-strand braid
with $N$ crossings, as shown in Fig. \ref{matrixbraid} and its
caption, keeping in mind that the state of an end-node is and can
only be with respect to its external edge. For the purpose of
calculation, it is also convenient to associate crossings with one
of the two end-nodes of a braid. For example, in Fig.
\ref{matrixbraid} the left end-node with its nearest crossing can be
denoted by $\oplus\left(
\begin{array}
[c]{c}%
-1\\
0
\end{array}
\right)  $, and the right end-node with its nearest two crossings
can be written as $\left(
\begin{array}
[c]{cc}%
0 & 0\\
+1 & +1
\end{array}
\right)  \ominus$, which has $\left(
\begin{array}
[c]{c}%
0\\
+1
\end{array}
\right)  \ominus$ as its 1-crossing \textbf{sub end-node}. End-nodes
represented in this way are called 1-crossing end-nodes, 2-crossing
end-nodes, etc. An end-node without any crossing is christened a
\textbf{bare end-node}.

A braid can be decomposed into or recombined from a left end-node, a
right
end-node, and a bunch of crossings represented by matrices. For instance,%
\[
\oplus\left(
\begin{array}
[c]{ccc}%
-1 & 0 & -1\\
0 & +1 & 0
\end{array}
\right)  \oplus\Longleftrightarrow\oplus\left(
\begin{array}
[c]{c}%
-1\\
0
\end{array}
\right)  +\left(
\begin{array}
[c]{c}%
+1\\
0
\end{array}
\right)  +\left(
\begin{array}
[c]{ccc}%
-1 & 0 & -1\\
0 & +1 & 0
\end{array}
\right)  \oplus,
\]
where the "$+$" between two matrices on the RHS means direct sum or
concatenation of two pieces of braids. One can see from the above
equation that the first two crossings or the second and third
crossings on the RHS are cancelled. Given this, we have the
following definition.

\begin{definition}
\label{defRedunode}An $N$-crossing end-node is said to be a
\textbf{reducible end-node}, if it is equivalent to a $M$-crossing
end-node with $M<N$, by equivalence moves done on the node;
otherwise, it is irreducible.
\end{definition}

The definition above gives rise to another definition of reducible
braid, which is equivalent to Definition \ref{defRedubraid}.

\begin{definition}
\label{defRedubraid2}A braid is said to be \textbf{reducible} if it
has a reducible end-node. If a braid has a reducible left or right
end-node, or both, it is called \textbf{left-}, or \textbf{right-,
}or\textbf{ two-way-reducible.}
\end{definition}

For consistency we may also symbolize the rotation moves. Because
rotations are exerted only on the end-nodes of a braid, we can
denote all possible moves by rotation operators $R_{LL}^{\theta}$,
$R_{LR}^{\theta}$, $R_{RL}^{\theta}$, and $R_{RR}^{\theta}$, where
the superscript $\theta$ is the angle of rotation, the first
subscript $L$ ($R$) reads that the operation is on the left (right)
end-node of a braid, and the second subscript $L$ ($R$) indicates
that the direction of rotation is left- (right-) handed. The left-
(right-) handedness of the rotation is defined in such a way that if
you grab the rotation axis of a node in your left (right) hand, with
the thumb pointing to the node, your hand wraps up in the direction
of rotation. Results of the rotation operators have been shown
graphically in Fig. \ref{pi3rot+} through Fig. \ref{pirot-}.
Here we show an example of the algebra in the following equation.%
\begin{align*}
&  R_{RR}^{\pi/3}\left[  \oplus\left(
\begin{array}
[c]{ccc}%
-1 & 0 & 0\\
0 & +1 & +1
\end{array}
\right)  \ominus\right]  \\
&  =\oplus\left(
\begin{array}
[c]{ccc}%
-1 & 0 & 0\\
0 & +1 & +1
\end{array}
\right)  R_{RR}^{\pi/3}\left(  \ominus\right)  \\
&  =\oplus\left(
\begin{array}
[c]{ccc}%
-1 & 0 & 0\\
0 & +1 & +1
\end{array}
\right)  +\left(
\begin{array}
[c]{c}%
-1\\
0
\end{array}
\right)  \oplus\\
&  =\oplus\left(
\begin{array}
[c]{cccc}%
-1 & 0 & 0 & -1\\
0 & +1 & +1 & 0
\end{array}
\right)  \oplus.
\end{align*}

Because a braid can be reduced only from its end-nodes, we first
classify the (ir)reducible end-nodes. We start from 1-crossing
end-nodes; all the possible ones are listed in table
\ref{tb:all1Xnodes}

\begin{table}[ptbh]
\begin{center}%
\begin{tabular}
[c]{|l|l|l|}\hline Left end-nodes &  & Right end-nodes\\\hline
\multicolumn{1}{|c|}{$\oplus\left(
\begin{array}
[c]{c}%
\pm1\\
0
\end{array}
\right)  $} & \multicolumn{1}{|c|}{} & \multicolumn{1}{|c|}{$\left(
\begin{array}
[c]{c}%
\pm1\\
0
\end{array}
\right)  \oplus$}\\\hline \multicolumn{1}{|c|}{$\oplus\left(
\begin{array}
[c]{c}%
0\\
\pm1
\end{array}
\right)  $} & \multicolumn{1}{|c|}{} & \multicolumn{1}{|c|}{$\left(
\begin{array}
[c]{c}%
0\\
\pm1
\end{array}
\right)  \oplus$}\\\hline \multicolumn{1}{|c|}{$\ominus\left(
\begin{array}
[c]{c}%
\pm1\\
0
\end{array}
\right)  $} & \multicolumn{1}{|c|}{} & \multicolumn{1}{|c|}{$\left(
\begin{array}
[c]{c}%
\pm1\\
0
\end{array}
\right)  \ominus$}\\\hline \multicolumn{1}{|c|}{$\ominus\left(
\begin{array}
[c]{c}%
0\\
\pm1
\end{array}
\right)  $} & \multicolumn{1}{|c|}{} & \multicolumn{1}{|c|}{$\left(
\begin{array}
[c]{c}%
0\\
\pm1
\end{array}
\right)  \ominus$}\\\hline
\end{tabular}
\end{center}
\caption{Table of all possible 1-crossing end-nodes.}%
\label{tb:all1Xnodes}%
\end{table}

The following equations then show how all the reducible 1-crossing
end-nodes
are reduced to bare nodes.%
\begin{align}
R_{LR}^{\pi/3}\left[  \oplus\left(
\begin{array}
[c]{c}%
+1\\
0
\end{array}
\right)  \right]   &  =\ominus\left(
\begin{array}
[c]{c}%
-1\\
0
\end{array}
\right)  +\left(
\begin{array}
[c]{c}%
+1\\
0
\end{array}
\right)  =\ominus\left(
\begin{array}
[c]{c}%
0\\
0
\end{array}
\right) \label{reduce1x}\\
R_{LL}^{\pi/3}\left[  \oplus\left(
\begin{array}
[c]{c}%
0\\
-1
\end{array}
\right)  \right]   &  =\ominus\left(
\begin{array}
[c]{c}%
0\\
+1
\end{array}
\right)  +\left(
\begin{array}
[c]{c}%
0\\
-1
\end{array}
\right)  =\ominus\left(
\begin{array}
[c]{c}%
0\\
0
\end{array}
\right) \nonumber\\
R_{LL}^{\pi/3}\left[  \ominus\left(
\begin{array}
[c]{c}%
-1\\
0
\end{array}
\right)  \right]   &  =\oplus\left(
\begin{array}
[c]{c}%
+1\\
0
\end{array}
\right)  +\left(
\begin{array}
[c]{c}%
-1\\
0
\end{array}
\right)  =\oplus\left(
\begin{array}
[c]{c}%
0\\
0
\end{array}
\right) \nonumber\\
R_{LR}^{\pi/3}\left[  \ominus\left(
\begin{array}
[c]{c}%
0\\
+1
\end{array}
\right)  \right]   &  =\oplus\left(
\begin{array}
[c]{c}%
0\\
-1
\end{array}
\right)  +\left(
\begin{array}
[c]{c}%
0\\
+1
\end{array}
\right)  =\oplus\left(
\begin{array}
[c]{c}%
0\\
0
\end{array}
\right) \nonumber\\
R_{RL}^{\pi/3}\left[  \left(
\begin{array}
[c]{c}%
-1\\
0
\end{array}
\right)  \oplus\right]   &  =\left(
\begin{array}
[c]{c}%
-1\\
0
\end{array}
\right)  +\left(
\begin{array}
[c]{c}%
+1\\
0
\end{array}
\right)  \ominus=\left(
\begin{array}
[c]{c}%
0\\
0
\end{array}
\right)  \ominus\nonumber\\
R_{RR}^{\pi/3}\left[  \left(
\begin{array}
[c]{c}%
0\\
+1
\end{array}
\right)  \oplus\right]   &  =\left(
\begin{array}
[c]{c}%
0\\
+1
\end{array}
\right)  +\left(
\begin{array}
[c]{c}%
0\\
-1
\end{array}
\right)  \ominus=\left(
\begin{array}
[c]{c}%
0\\
0
\end{array}
\right)  \ominus\nonumber\\
R_{RR}^{\pi/3}\left[  \left(
\begin{array}
[c]{c}%
+1\\
0
\end{array}
\right)  \ominus\right]   &  =\left(
\begin{array}
[c]{c}%
+1\\
0
\end{array}
\right)  +\left(
\begin{array}
[c]{c}%
-1\\
0
\end{array}
\right)  \oplus=\left(
\begin{array}
[c]{c}%
0\\
0
\end{array}
\right)  \oplus\nonumber\\
R_{RL}^{\pi/3}\left[  \left(
\begin{array}
[c]{c}%
0\\
-1
\end{array}
\right)  \ominus\right]   &  =\left(
\begin{array}
[c]{c}%
0\\
-1
\end{array}
\right)  +\left(
\begin{array}
[c]{c}%
0\\
+1
\end{array}
\right)  \oplus=\left(
\begin{array}
[c]{c}%
0\\
0
\end{array}
\right)  \oplus\nonumber
\end{align}
With the help of the above calculations, we can easily list all the
irreducible 1-crossing end-nodes in table \ref{tb:irred1Xnodes}.%
\begin{table}[ptbh]
\begin{center}%
\begin{tabular}
[c]{|l|l|l|}\hline Left end-nodes &  & Right end-nodes\\\hline
\multicolumn{1}{|c|}{$\oplus\left(
\begin{array}
[c]{c}%
-1\\
0
\end{array}
\right)  $} & \multicolumn{1}{|c|}{} & \multicolumn{1}{|c|}{$\left(
\begin{array}
[c]{c}%
+1\\
0
\end{array}
\right)  \oplus$}\\\hline \multicolumn{1}{|c|}{$\oplus\left(
\begin{array}
[c]{c}%
0\\
+1
\end{array}
\right)  $} & \multicolumn{1}{|c|}{} & \multicolumn{1}{|c|}{$\left(
\begin{array}
[c]{c}%
0\\
-1
\end{array}
\right)  \oplus$}\\\hline \multicolumn{1}{|c|}{$\ominus\left(
\begin{array}
[c]{c}%
+1\\
0
\end{array}
\right)  $} & \multicolumn{1}{|c|}{} & \multicolumn{1}{|c|}{$\left(
\begin{array}
[c]{c}%
-1\\
0
\end{array}
\right)  \ominus$}\\\hline \multicolumn{1}{|c|}{$\ominus\left(
\begin{array}
[c]{c}%
0\\
-1
\end{array}
\right)  $} & \multicolumn{1}{|c|}{} & \multicolumn{1}{|c|}{$\left(
\begin{array}
[c]{c}%
0\\
+1
\end{array}
\right)  \ominus$}\\\hline
\end{tabular}
\end{center}
\caption{Table of irreducible 1-crossing end-nodes.}%
\label{tb:irred1Xnodes}%
\end{table}

Now we consider 2-crossing end-nodes; there are a total of 48 of
this kind, including left and right end-nodes. To find all the
irreducible 2-crossing end-nodes, we need only to think about those
whose sub 1-crossing nodes are irreducible, since otherwise a
2-crossing end-node is already reducible; this excludes 24
2-crossing end-nodes. If a 2-crossing node has an irreducible sub
1-crossing node, its crossings can definitely not be reduced by
$2\pi /3$-rotations, because a $2\pi/3$-rotation is made of two
consecutive $\pi /3$-rotations that do not reduce any irreducible
1-crossing node, and it does not flip the state of a bare node.
Interestingly, however, a 2-crossing end-node with an irreducible
sub 1-crossing end-node may still be reduced to an irreducible
1-crossing end-node by $\pi$-rotations, which can be seen from the
following equations.%
\begin{align}
R_{LL}^{\pi}\left[  \oplus\left(
\begin{array}
[c]{cc}%
-1 & 0\\
0 & -1
\end{array}
\right)  \right]   &  =\ominus\left(
\begin{array}
[c]{ccc}%
+1 & 0 & +1\\
0 & +1 & 0
\end{array}
\right)  +\left(
\begin{array}
[c]{cc}%
-1 & 0\\
0 & -1
\end{array}
\right)  =\ominus\left(
\begin{array}
[c]{c}%
+1\\
0
\end{array}
\right)  \label{reduce2x}\\
R_{LR}^{\pi}\left[  \oplus\left(
\begin{array}
[c]{cc}%
0 & +1\\
+1 & 0
\end{array}
\right)  \right]   &  =\ominus\left(
\begin{array}
[c]{ccc}%
0 & -1 & 0\\
-1 & 0 & -1
\end{array}
\right)  +\left(
\begin{array}
[c]{cc}%
0 & +1\\
+1 & 0
\end{array}
\right)  =\ominus\left(
\begin{array}
[c]{c}%
0\\
-1
\end{array}
\right)  \nonumber\\
R_{LR}^{\pi}\left[  \ominus\left(
\begin{array}
[c]{cc}%
+1 & 0\\
0 & +1
\end{array}
\right)  \right]   &  =\oplus\left(
\begin{array}
[c]{ccc}%
-1 & 0 & -1\\
0 & -1 & 0
\end{array}
\right)  +\left(
\begin{array}
[c]{cc}%
+1 & 0\\
0 & +1
\end{array}
\right)  =\oplus\left(
\begin{array}
[c]{c}%
-1\\
0
\end{array}
\right)  \nonumber\\
R_{LL}^{\pi}\left[  \ominus\left(
\begin{array}
[c]{cc}%
0 & -1\\
-1 & 0
\end{array}
\right)  \right]   &  =\oplus\left(
\begin{array}
[c]{ccc}%
0 & +1 & 0\\
+1 & 0 & +1
\end{array}
\right)  +\left(
\begin{array}
[c]{cc}%
0 & -1\\
-1 & 0
\end{array}
\right)  =\oplus\left(
\begin{array}
[c]{c}%
0\\
+1
\end{array}
\right)  \nonumber\\
R_{RR}^{\pi}\left[  \left(
\begin{array}
[c]{cc}%
0 & +1\\
+1 & 0
\end{array}
\right)  \oplus\right]   &  =\left(
\begin{array}
[c]{cc}%
0 & +1\\
+1 & 0
\end{array}
\right)  +\left(
\begin{array}
[c]{ccc}%
-1 & 0 & -1\\
0 & -1 & 0
\end{array}
\right)  \ominus=\left(
\begin{array}
[c]{c}%
-1\\
0
\end{array}
\right)  \ominus\nonumber\\
R_{RL}^{\pi}\left[  \left(
\begin{array}
[c]{cc}%
-1 & 0\\
0 & -1
\end{array}
\right)  \oplus\right]   &  =\left(
\begin{array}
[c]{cc}%
-1 & 0\\
0 & -1
\end{array}
\right)  +\left(
\begin{array}
[c]{ccc}%
0 & +1 & 0\\
+1 & 0 & +1
\end{array}
\right)  \ominus=\left(
\begin{array}
[c]{c}%
0\\
+1
\end{array}
\right)  \ominus\nonumber\\
R_{LL}^{\pi}\left[  \left(
\begin{array}
[c]{cc}%
0 & -1\\
-1 & 0
\end{array}
\right)  \ominus\right]   &  =\left(
\begin{array}
[c]{cc}%
0 & -1\\
-1 & 0
\end{array}
\right)  +\left(
\begin{array}
[c]{ccc}%
+1 & 0 & +1\\
0 & +1 & 0
\end{array}
\right)  \oplus=\left(
\begin{array}
[c]{c}%
+1\\
0
\end{array}
\right)  \oplus\nonumber\\
R_{LR}^{\pi}\left[  \left(
\begin{array}
[c]{cc}%
+1 & 0\\
0 & +1
\end{array}
\right)  \ominus\right]   &  =\left(
\begin{array}
[c]{cc}%
+1 & 0\\
0 & +1
\end{array}
\right)  +\left(
\begin{array}
[c]{ccc}%
0 & -1 & 0\\
-1 & 0 & -1
\end{array}
\right)  \oplus=\left(
\begin{array}
[c]{c}%
0\\
-1
\end{array}
\right)  \oplus\nonumber
\end{align}
Consequently, we can list all the irreducible 2-crossing end-nodes
in table \ref{tb:irred2Xnodes}
\begin{table}[ptbh]
\begin{center}%
\begin{tabular}
[c]{|c|l|c|c|l|}\hline \multicolumn{2}{|c|}{Left end-nodes} &  &
\multicolumn{2}{|c|}{Right end-nodes}\\\hline
\multicolumn{1}{|c|}{$\oplus\left(
\begin{array}
[c]{cc}%
-1 & 0\\
0 & +1
\end{array}
\right)  $} & $\oplus\left(
\begin{array}
[c]{cc}%
-1 & -1\\
0 & 0
\end{array}
\right)  $ & \multicolumn{1}{|c|}{} & $\left(
\begin{array}
[c]{cc}%
0 & +1\\
-1 & 0
\end{array}
\right)  \oplus$ & $\left(
\begin{array}
[c]{cc}%
+1 & +1\\
0 & 0
\end{array}
\right)  \oplus$\\\hline \multicolumn{1}{|c|}{$\oplus\left(
\begin{array}
[c]{cc}%
0 & -1\\
+1 & 0
\end{array}
\right)  $} & $\oplus\left(
\begin{array}
[c]{cc}%
0 & 0\\
+1 & +1
\end{array}
\right)  $ & \multicolumn{1}{|c|}{} & $\left(
\begin{array}
[c]{cc}%
+1 & 0\\
0 & -1
\end{array}
\right)  \oplus$ & $\left(
\begin{array}
[c]{cc}%
0 & 0\\
-1 & -1
\end{array}
\right)  \oplus$\\\hline \multicolumn{1}{|c|}{$\ominus\left(
\begin{array}
[c]{cc}%
+1 & 0\\
0 & -1
\end{array}
\right)  $} & $\ominus\left(
\begin{array}
[c]{cc}%
+1 & +1\\
0 & 0
\end{array}
\right)  $ & \multicolumn{1}{|c|}{} & $\left(
\begin{array}
[c]{cc}%
0 & -1\\
+1 & 0
\end{array}
\right)  \ominus$ & $\left(
\begin{array}
[c]{cc}%
-1 & -1\\
0 & 0
\end{array}
\right)  \ominus$\\\hline \multicolumn{1}{|c|}{$\ominus\left(
\begin{array}
[c]{cc}%
0 & +1\\
-1 & 0
\end{array}
\right)  $} & $\ominus\left(
\begin{array}
[c]{cc}%
0 & 0\\
-1 & -1
\end{array}
\right)  $ & \multicolumn{1}{|c|}{} & $\left(
\begin{array}
[c]{cc}%
-1 & 0\\
0 & +1
\end{array}
\right)  \ominus$ & $\left(
\begin{array}
[c]{cc}%
0 & 0\\
+1 & +1
\end{array}
\right)  \ominus$\\\hline
\end{tabular}
\end{center}
\caption{Table of irreducible 2-crossing end-nodes.}%
\label{tb:irred2Xnodes}%
\end{table}

The following theorem states that there is no need to investigate
end-nodes with more crossings to see if they are irreducible.

\begin{theorem}
\label{theoNXnode}An $N$-crossing end-node, $N>2$, which has an
irreducible 2-crossing sub end-node, is irreducible.
\end{theorem}

\begin{proof}
If the $N$-crossing end-node has a irreducible 2-crossing sub
end-node, the two crossings nearest to the node are not reducible by
either a single $\pi /3$- or a single $2\pi/3$-rotation on the node.
We may consider $\pi $-rotations on its 3-crossing sub end-node.
However, if a 3-crossing end-node is reducible by a $\pi$-rotation,
it must contain a reducible 2-crossing sub end-node according to
Fig. \ref{pirot+}, Fig. \ref{pirot-} and Eq.\ref{reduce2x}, which is
contradictory to the condition given in the theorem. This is then
true for all cases where $N>3$ by simple induction. Therefore, the
theorem holds.
\end{proof}

Equipped with the knowledge of (ir)reducible end-nodes, we are ready
to classify braids. The two end-nodes of a braid are either in the
same states or in opposite states, we first take a look at braids
whose end nodes are in the same states.

\begin{theorem}
\label{theo123Xbraids}All $N$-crossing braids in the form
$\oplus\left(
\begin{array}
[c]{ccc}
& \cdots & \\
& \cdots &
\end{array}
\right)  \oplus$ and $\ominus\left(
\begin{array}
[c]{ccc}
& \cdots & \\
& \cdots &
\end{array}
\right)  \ominus$ are reducible for $N\leq3$.
\end{theorem}

\begin{proof}
It suffices to prove the $\oplus\oplus$ case, the case of
$\ominus\ominus$ follows similarly or by symmetry.

1) $N=1$. There are only four possibilities, namely $\oplus\left(
\begin{array}
[c]{c}%
\pm1\\
0
\end{array}
\right)  \oplus$ and $\oplus\left(
\begin{array}
[c]{c}%
0\\
\pm1
\end{array}
\right)  \oplus$; however, they are all reducible because they all
contain one reducible 1-crossing end-node according to Eq.
\ref{reduce1x}.

2) $N=2$. We first consider the braids formed by an irreducible
2-crossing end-node $\oplus_{irred}\left(
2\text{\textrm{-crossing}}\right)  $ or $\left(
2\text{\textrm{-crossing}}\right)  \oplus_{irred}$, and a bare
end-node $\oplus_{b}$. We do the following decomposition%
\begin{align*}
\oplus_{irred}\left(  2\text{\textrm{-crossing}}\right)  +\oplus_{b}
&
=\oplus_{irred}\left(  2\text{\textrm{-crossing}}\right)  \oplus_{b}%
=\oplus_{irred}+\left(  2\text{\textrm{-crossing}}\right)  \oplus_{b}\\
\oplus_{b}+\left(  2\text{\textrm{-crossing}}\right)  \oplus_{irred}
&
=\oplus_{b}\left(  2\text{\textrm{-crossing}}\right)  \oplus_{irred}%
=\oplus_{b}\left(  2\text{\textrm{-crossing}}\right)
+\oplus_{irred}.
\end{align*}
Then from table \ref{tb:irred2Xnodes}, it is readily seen that
$\left( 2\text{\textrm{-crossing}}\right)  \oplus_{b}$ and
$\oplus_{b}\left( 2\text{\textrm{-crossing}}\right)  $ are always
reducible end-nodes for any choice of $\oplus_{irred}\left(
2\text{\textrm{-crossing}}\right)  $ and $\left(
2\text{\textrm{-crossing}}\right)  \oplus_{irred}$respectively. That
is, the braids formed this way are reducible. We then consider
braids formed by two irreducible 1-crossing end-nodes. The first two
rows in table \ref{tb:irred1Xnodes} and Eq. \ref{reduce2x} clearly
shows that the result is either an unbraid or one with a reducible
2-crossing end-node.

3) $N=3$. We need only consider braids, each of which is formed by
the direct sum of a 2-crossing irreducible end-node and a 1-crossing
irreducible end-node. This can be done by taking the direct sum
between the (right) left end-nodes in the first two rows of table
\ref{tb:irred1Xnodes}, and the (left) right end-nodes in the first
two rows of table \ref{tb:irred2Xnodes}. It is not hard to see that
any resultant braid has merely two possibilities: i) two neighboring
crossings are cancelled by the direct sum, which leads to 1-crossing
braids that are proven to be reducible in the case of $N=1$; and ii)
a crossing in the irreducible 2-crossing end-node is combined with
the irreducible 1-crossing end-node to form a reducible 2-crossing
end-node, i.e. the braid is reducible.
\end{proof}

Theorem \ref{theo123Xbraids} does not cover the case where $N\geq4$,
which will be included in another theorem soon. Before that, let us
consider the braids whose end-nodes are in opposite states.

\begin{case}
$N$-crossing braids in the form $\oplus\left(
\begin{array}
[c]{ccc}
& \cdots & \\
& \cdots &
\end{array}
\right)  \ominus$ and $\ominus\left(
\begin{array}
[c]{ccc}
& \cdots & \\
& \cdots &
\end{array}
\right)  \oplus$, for $N\leq3$. Note that due to
Theorem\ref{theo123Xbraids}, the set of irreducible 1-crossing
braids to be found here represents the full set of irreducible
braids for $N\leq3$, regardless of the states of the end-nodes.
\end{case}

\begin{enumerate}
\item $N=1$. An irreducible braid can only be made by an irreducible
1-crossing end-node and a bare node. From table
\ref{tb:irred1Xnodes}, there are only four options, which are indeed
all irreducible; they are now listed in table
\ref{tb:irred1Xbraids}. \begin{table}[ptbh]
\begin{center}%
\begin{tabular}
[c]{|l|l|l|}\hline \multicolumn{1}{|c|}{$\oplus\left(
\begin{array}
[c]{c}%
-1\\
0
\end{array}
\right)  \ominus$} & \multicolumn{1}{|c|}{} &
\multicolumn{1}{|c|}{$\ominus \left(
\begin{array}
[c]{c}%
+1\\
0
\end{array}
\right)  \oplus$}\\\hline \multicolumn{1}{|c|}{$\oplus\left(
\begin{array}
[c]{c}%
0\\
+1
\end{array}
\right)  \ominus$} & \multicolumn{1}{|c|}{} &
\multicolumn{1}{|c|}{$\ominus \left(
\begin{array}
[c]{c}%
0\\
-1
\end{array}
\right)  \oplus$}\\\hline
\end{tabular}
\end{center}
\caption{Table of irreducible 1-crossing braids.}%
\label{tb:irred1Xbraids}%
\end{table}

\item $N=2$. It is sufficient to consider the braids formed by an irreducible
2-crossing end-node and a bare end-node in the opposite state. The
reason is that if a 2-crossing braid is irreducible, its two
2-crossing end-nodes must be irreducible as well; moreover, if a
2-crossing end-node is irreducible, its 1-crossing sub end-node is
already irreducible. Therefore, one can simply add to each
irreducible end-node in table \ref{tb:irred2Xnodes} a bare end-node
in the opposite state to create an irreducible 2-crossing braid.
Being a bit redundant, we list all the 16 irreducible 2-crossing
braids in table \ref{tb:irred2Xbraids}.\begin{table}[ptbh]
\begin{center}%
\begin{tabular}
[c]{|c|l|c|c|l|}\hline \multicolumn{1}{|c|}{$\oplus\left(
\begin{array}
[c]{cc}%
-1 & 0\\
0 & +1
\end{array}
\right)  \ominus$} & $\oplus\left(
\begin{array}
[c]{cc}%
-1 & -1\\
0 & 0
\end{array}
\right)  \ominus$ & \multicolumn{1}{|c|}{} & $\ominus\left(
\begin{array}
[c]{cc}%
0 & +1\\
-1 & 0
\end{array}
\right)  \oplus$ & $\ominus\left(
\begin{array}
[c]{cc}%
+1 & +1\\
0 & 0
\end{array}
\right)  \oplus$\\\hline \multicolumn{1}{|c|}{$\oplus\left(
\begin{array}
[c]{cc}%
0 & -1\\
+1 & 0
\end{array}
\right)  \ominus$} & $\oplus\left(
\begin{array}
[c]{cc}%
0 & 0\\
+1 & +1
\end{array}
\right)  \ominus$ & \multicolumn{1}{|c|}{} & $\ominus\left(
\begin{array}
[c]{cc}%
+1 & 0\\
0 & -1
\end{array}
\right)  \oplus$ & $\ominus\left(
\begin{array}
[c]{cc}%
0 & 0\\
-1 & -1
\end{array}
\right)  \oplus$\\\hline \multicolumn{1}{|c|}{$\ominus\left(
\begin{array}
[c]{cc}%
+1 & 0\\
0 & -1
\end{array}
\right)  \oplus$} & $\ominus\left(
\begin{array}
[c]{cc}%
+1 & +1\\
0 & 0
\end{array}
\right)  \oplus$ & \multicolumn{1}{|c|}{} & $\oplus\left(
\begin{array}
[c]{cc}%
0 & -1\\
+1 & 0
\end{array}
\right)  \ominus$ & $\oplus\left(
\begin{array}
[c]{cc}%
-1 & -1\\
0 & 0
\end{array}
\right)  \ominus$\\\hline \multicolumn{1}{|c|}{$\ominus\left(
\begin{array}
[c]{cc}%
0 & +1\\
-1 & 0
\end{array}
\right)  \oplus$} & $\ominus\left(
\begin{array}
[c]{cc}%
0 & 0\\
-1 & -1
\end{array}
\right)  \oplus$ & \multicolumn{1}{|c|}{} & $\oplus\left(
\begin{array}
[c]{cc}%
-1 & 0\\
0 & +1
\end{array}
\right)  \ominus$ & $\oplus\left(
\begin{array}
[c]{cc}%
0 & 0\\
+1 & +1
\end{array}
\right)  \ominus$\\\hline
\end{tabular}
\end{center}
\caption{Table of irreducible 2-crossing braids.}%
\label{tb:irred2Xbraids}%
\end{table}

\item $N=3$. A 3-crossing braid in this case is irreducible if and only if it
admits the following two decompositions.%
\begin{align*}
&  \text{An irreducible 1-crossing end-node }+\text{ An irreducible
2-crossing
end-node}\\
&  \text{An irreducible 2-crossing end-node }+\text{ An irreducible
1-crossing
end-node,}%
\end{align*}
where "$+$" is understood as the direct sum. The proof of this claim
follows immediately from Theorem \ref{theoNXnode}.
\end{enumerate}

It is time to summarize the case of $N\geq4$ for $N$-crossing
braids, regardless of the states of the end-nodes, by the following
theorem.

\begin{theorem}
\label{theoNXbraids}A $N$-crossing braid for $N\geq4$ is
irreducible, if and
only if it admits the decomposition%
\begin{align*}
&  \text{An irreducible 2-crossing end-node }\\
+\text{ } &  \text{Arbitrary sequence of crossings }\\
+\text{ } &  \text{An irreducible 2-crossing end-node,}%
\end{align*}
where "$+$" is understood as the direct sum. The only constraint of
the arbitrary sequence of crossings is that its last crossing on
each side does not cancel the neighboring crossing associated with
the end-node on the same side.
\end{theorem}

\begin{proof}
An irreducible 2-crossing end-node contains an irreducible
1-crossing end-node. By theorem \ref{theoNXnode}, if the above
decomposition is admitted, the braid is not reducible on either
end-node whatever the arbitrary sequence of crossings is up to the
constraint. Therefore, the theorem holds.
\end{proof}

The braids that are interesting to us are those reducible ones,
which is shown in the companion paper. Thus we may make more
detailed divisions in the type of reducible braids by the definition
below.

\begin{definition}
Given a reducible braid $B$, a braid $B^{\prime}$ obtained from $B$
by doing as much reduction as possible is called an
\textbf{extremum} of $B$; $B$ may have more than one
\textbf{extrema}, but all the extrema have the same number of
crossings. We then have the following.

\begin{enumerate}
\item If all extrema of $B$ are unbraids, i.e. braids with no crossing, $B$ is
said to be \textbf{completely reducible.}

\item if an extremum of $B$ can be reached by equivalence moves exerted only
on its (left)right end-node, $B$ is called \textbf{extremely}
(\textbf{left-)right-reducible;} if $B$ is also completely
reducible, $B$ is then said to be \textbf{completely
(left-)right-reducible.} Note that completely (left-)
right-reducible implies extremely (left-) right-reducible, but not
vice versa in general.
\end{enumerate}
\end{definition}

\bigskip

\section{Conclusions \& Perspectives}

In this paper, we proposed a new notation, namely the tube-sphere
notation, for embedded (framed) 4-valent spin-networks. By means of
this notation, we discovered a type of topological structures, the
3-strand braids, as sub-diagrams of an embedded spin-net.
Equivalence moves, including translations and rotations, which
divide projections of embeddings of spin-networks into different
equivalence classes, are defined and discussed in detail. The
equivalence moves are important and useful in two aspects. Firstly,
by rotations, we classify 3-strand braids into two major types:
reducible braids and irreducible braids, the former of which are
further classified for the purpose of subsequent works. Secondly, by
equivalence moves one is able to carry out the calculation of braid
propagation and interactions of embedded 4-valent spin-nets.

These results serve as foundations for the work in the companion
paper and all our future work dealing with braid-like excitations of
embedded 4-valent spin-networks. In another paper, we will propose
the evolution moves of embedded 4-valent spin-networks, by which
some of the (reducible) 3-strand braids are able to propagate on the
spin-nets and interact with each other and provide a possible
formulation of the dynamics of these local excitations.

\bigskip

\section*{Acknowledgements}
The author is in debt to Lee Smolin, the author's advisor, for his
great insight and heuristic discussion. He is grateful to Fotini
Markopoulou for her critical comments. He would appreciate the
helpful discussions with Isabeau Premont-Schwarz, Aristide Baratin,
and Thomasz Konopka. Gratitude must also go to Sundance
Bilson-Thompson for his proof-reading of the manuscript. Research at
Perimeter Institute is supported in part by the Government of Canada
through NSERC and by the Province of Ontario through MEDT.

\newif\ifabfull\abfulltrue%

\end{document}